# FDI, banking crisis and growth: direct and spill over effects

*Brahim GAIES*
*IPAG Lab - IPAG Business School, Paris, France*

*Khaled GUESMI*
*IPAG Lab - IPAG Business School, Paris, France*

*Stéphane GOUTTE*
*Paris School of Business, Paris, France*

**ABSTRACT**
This study suggests a new decomposition of the effect of Foreign Direct Investment (FDI) on long-term growth in developing countries. It reveals that FDI not only have a positive direct effect on growth, but also increase the latter by reducing the recessionary effect resulting from a banking crisis. Even more, they reduce its occurrence.

## I. Introduction

How do FDI impact economic growth in developing countries? According to recent studies on the effects of financial globalization, the answer to this question is that the spill over benefits of FDI are more important than their direct advantages. In this vein, Okada (2013) shows a positive interaction effect between FDI and institutional quality on GDP. He concludes that FDI directly increase growth and indirectly enhance it by promoting institutional quality. Meanwhile, Neto and Veiga (2013) find the same result for the triplet FDI, technology and growth. On the other hand, Kunieda, Okada and Shibata (2014) highlight that financial development is one of the major spill over benefits of FDI on growth. The same conclusion is drawn by Ahmed (2016) and Trabelsi and Cherif (2017). In line with these investigations, Iamsiraroj (2016) presents empirical evidence concerning a spill over benefit of FDI on growth, which passes through the human capital quality channel. Regarding this literature, the main originality of this study is that it presents a new decomposition of the effect (direct and spill over) of FDI on growth, by considering their interaction with banking crisis, in 67 developing countries, among low and lower-middle income countries according to the World Bank classification, between 1972 and 2011.

## II. Methodology

To decompose the effect of FDI on economic growth, we specify three models, namely two economic growth models (Equations (1) and (2)) and one banking crisis model (Equation (3)).

$$GDPPCG_{it} = \alpha_0 + \gamma\, GDPPC_{it-1} + \alpha_1\, FDI_{it} + \alpha_2 CRISIS + + \beta X_{it} + \mu_i + \lambda_t + \varepsilon_{it} \qquad (1)$$

$$GDPPCG_{it} = \alpha_0 + \gamma\, GDPPC_{it-1} + \alpha_1\, FDI_{it} + \alpha_2 CRISIS + \alpha_3\, (FDI_{it} \times CRISIS) + \beta X_{it} + \mu_i + \lambda_t + \varepsilon_{it} \qquad (2)$$

In Equations (1) and (2) above, GDPPCG represents the dependent variable, namely real GDP per capita growth. FDI is the first interest variables. It is measured by the total stocks of FDI assets and liabilities to GDP (FDIT) or the stocks of FDI liabilities to GDP (FDIL), extracted from the External Wealth of Nations Dataset. CRISIS is the second interest variable. It is taken from the Systemic Banking Crises Database (IMF, last update 2012). It equals "one" if there is a banking crisis in country *i* in year *t*, and "zero" otherwise. The control variables are GDPPC, which is the real GDP per capita lagged and X, which is a matrix regrouping the sum of exports and imports to GDP (TRADE); terms-of-trade growth (TERM); government spending to GDP (GOV); and secondary school enrolment (EDU). These variables are obtained from the World Bank Indicators database (WDI). $\alpha_0$ is a constant; $\mu_i$ is the country-specific effect; $\lambda t$ is the time-specific effect; and $\varepsilon_{it}$ is the error term. Also, to estimate these growth models and deal with endogeneity bias (Roodman, 2009), we utilize the GMM system dynamic panel data estimator (Two-step system GMM) developed by Arellano and Boyer (1995) and Blundell and Bond (1998), and we compute robust two-step standard errors by following the methodology proposed by Newey and Windmeijer (2009). In addition, we use the least square dummy variables estimator, the random-effects estimator and the one-step system GMM estimator to test the robustness of the Two-step system GMM results.

$$CRISIS_{it} = F(\Omega_1 FDI_{it} + \beta Z_{it} + \alpha_0 + \varepsilon_{it}) \qquad (3)$$

In Equation (3) above, CRISIS is explained by FDI variables and Z matrix, which represents the set of control variables. These are GDPPC, GOV, the GDP growth (GROWTH), the growth of money and quasi money to total reserves ratio (M2toRES), the growth of claims on private sector to GDP (CLAIM-PRIV), the domestic credit to private sector (% of GDP) (CPRIVET) and the inflation rate (INF), extracted from WDI, as well as POLI, which is the indicator of political rights (1 = most free and 7 = least free) obtained from the Freedom House database. $F(.)$ is the standard normal cumulative distribution function. Also, to estimate the banking crisis model and test the robustness of our results, we use three logit panel models (conditional fixed-effects, random-effects and population-averaged).

### III. Results and interpretations

Tables 1 and 2 below show that the coefficients of the variables FDIT and FDIL are significant and positive, and those of CRISIS are significant and negative in all regressions. This proves the positive direct effect of FDI on growth, and the negative direct effect of banking crisis on the latter. In addition, the coefficients of the interaction terms FDIT x CRISIS and FDIL x CRISIS are significant and positive in all regressions. This result indicates that FDI also allow a spill over (indirect) benefit on growth in developing countries. It consists of decreasing the recessionary effect resulting from a banking crisis. Besides, the outputs of Table 3 strongly consolidate this result. Indeed, the negativity and significance of the marginal effect of FDIT and FDIL on CRISIS in all regressions evidence that FDI are a negative determinant of banking crisis occurrence. In sum, theses direct and spill over advantages can be theoretically explained by two mechanisms. First, FDI reduce the negative effect of banking crises because they promote supervision and risk managing in the domestic financial markets through the presence of foreign investors, which indirectly oblige the local institutions to improve their governance's quality under the fear of the "sudden stop". Second, foreign investors inherence liquidity and technology transfer, which catalyse the domestic industry and foreign trade. The latter, as well as domestic investment, which increases through FDI due to their complementarity (crowding effect), are the main engines of growth.

## Table 1. Growth – Baseline and robustness estimations

| Estimator | Two-step system GMM | | | | LSDV | | | |
|---|---|---|---|---|---|---|---|---|
| L.GDPPC | -0.033 | -0.033* | -0.031 | -0.031 | -0.070*** | -0.069*** | -0.070*** | -0.069*** |
| | (0.021) | (0.018) | (0.023) | (0.019) | (0.016) | (0.016) | (0.016) | (0.016) |
| **FDIT** | **0.049**** | **0.040**** | | | **0.041**** | **0.032*** | | |
| | (0.023) | (0.020) | | | (0.018) | (0.019) | | |
| **FDIL** | | | **0.050**** | **0.041**** | | | **0.042**** | **0.032*** |
| | | | (0.022) | (0.019) | | | (0.018) | (0.018) |
| **CRISIS** | **-0.027**** | **-0.042***** | **-0.027**** | **-0.043***** | **-0.017**** | **-0.027***** | **-0.017**** | **-0.028***** |
| | (0.013) | (0.014) | (0.013) | (0.012) | (0.008) | (0.010) | (0.008) | (0.010) |
| **FDIT X CRISIS** | | **0.113***** | | | | **0.068**** | | |
| | | (0.040) | | | | (0.034) | | |
| **FDIL X CRISIS** | | | | **0.122***** | | | | **0.071**** |
| | | | | (0.042) | | | | (0.034) |
| TRADE | 0.040*** | 0.041*** | 0.037*** | 0.039*** | 0.025*** | 0.025*** | 0.024*** | 0.025*** |
| | (0.012) | (0.012) | (0.014) | (0.013) | (0.007) | (0.007) | (0.007) | (0.007) |
| EDU | 0.016 | 0.014 | 0.017 | 0.013 | 0.017*** | 0.017*** | 0.017*** | 0.017*** |
| | (0.011) | (0.011) | (0.012) | (0.012) | (0.005) | (0.005) | (0.005) | (0.005) |
| TERM | -0.000 | -0.000 | 0.001 | 0.001 | 0.010* | 0.009 | 0.010* | 0.009 |
| | (0.018) | (0.014) | (0.018) | (0.014) | (0.006) | (0.006) | (0.006) | (0.006) |
| GOV | -0.014 | -0.012 | -0.016 | -0.013 | -0.018*** | -0.018*** | -0.018*** | -0.018*** |
| | (0.018) | (0.015) | (0.018) | (0.016) | (0.007) | (0.007) | (0.007) | (0.007) |
| Constant | 0.037 | 0.034 | 0.038 | 0.033 | 0.295*** | 0.295*** | 0.295*** | 0.295*** |
| | (0.122) | (0.105) | (0.130) | (0.111) | (0.100) | (0.101) | (0.100) | (0.101) |
| *Observations* | 289 | 289 | 289 | 289 | 289 | 289 | 289 | 289 |
| *R2* | | | | | 0.685 | 0.689 | 0.685 | 0.690 |
| *AR2 P-value* | 0.178 | 0.139 | 0.186 | 0.138 | | | | |
| *Hansen P-value* | 0.466 | 0.616 | 0.436 | 0.585 | | | | |
| *Fisher* | | | | | 12.45 | 11.72 | 12.55 | 12.09 |

Non-overlapping five-year data. Standard errors are presented below the corresponding coefficient. Symbols *, ** and *** means significant at 10%, 5% and at 1%, respectively.

## Table 2. Growth – Other robustness estimations

| Estimator | One-step system GMM | | | | Random-effects | | | |
|---|---|---|---|---|---|---|---|---|
| L.GDPPC | -0.047** | -0.046** | -0.046** | -0.044** | -0.022*** | -0.022*** | -0.022*** | -0.022*** |
| | (0.020) | (0.021) | (0.022) | (0.020) | (0.008) | (0.008) | (0.008) | (0.008) |
| **FDIT** | **0.046**** | **0.034*** | | | **0.067***** | **0.057***** | | |
| | (0.020) | (0.020) | | | (0.015) | (0.014) | | |
| **FDIL** | | | **0.051**** | **0.034*** | | | **0.067***** | **0.056***** |
| | | | (0.021) | (0.018) | | | (0.015) | (0.014) |
| **CRISIS** | **-0.021*** | **-0.032**** | **-0.022*** | **-0.032**** | **-0.023**** | **-0.034***** | **-0.023**** | **-0.034***** |
| | (0.012) | (0.015) | (0.012) | (0.013) | (0.009) | (0.012) | (0.009) | (0.012) |
| **FDIT X CRISIS** | | **0.082*** | | | | **0.075**** | | |
| | | (0.044) | | | | (0.030) | | |
| **FDIL X CRISIS** | | | | **0.093**** | | | | **0.080***** |
| | | | | (0.045) | | | | (0.030) |
| TRADE | 0.033** | 0.036*** | 0.029** | 0.033*** | 0.021*** | 0.021*** | 0.021*** | 0.021*** |
| | (0.013) | (0.012) | (0.013) | (0.011) | (0.004) | (0.004) | (0.004) | (0.004) |
| EDU | 0.023** | 0.024** | 0.024** | 0.024** | 0.016*** | 0.016*** | 0.016*** | 0.016*** |
| | (0.009) | (0.010) | (0.010) | (0.009) | (0.004) | (0.004) | (0.004) | (0.004) |
| TERM | 0.028 | 0.022 | 0.029 | 0.014 | 0.006 | 0.005 | 0.006 | 0.005 |
| | (0.021) | (0.022) | (0.022) | (0.018) | (0.005) | (0.005) | (0.005) | (0.005) |
| GOV | -0.017 | -0.016 | -0.018 | -0.016 | -0.026*** | -0.026*** | -0.026*** | -0.026*** |
| | (0.015) | (0.015) | (0.016) | (0.014) | (0.005) | (0.005) | (0.005) | (0.005) |
| Constant | 0.007 | 0.010 | 0.008 | 0.054 | 0.043 | 0.049 | 0.041 | 0.048 |
| | (0.096) | (0.102) | (0.100) | (0.100) | (0.049) | (0.049) | (0.049) | (0.049) |
| *Observations* | 289 | 289 | 289 | 289 | 289 | 289 | 289 | 289 |
| *AR2 P-value* | 0.166 | 0.100 | 0.166 | 0.132 | | | | |
| *Hansen P-value* | 0.466 | 0.450 | 0.332 | 0.264 | | | | |
| *R2* | | | | | 0.487 | 0.483 | 0.487 | 0.482 |
| *Chi2-statistic* | | | | | 173.8 | 265.5 | 173.5 | 266.8 |

Non-overlapping five-year data. Standard errors are presented below the corresponding coefficient. Symbols *, ** and *** means significant at 10%, 5% and at 1%, respectively.

**Table 3. Crisis– Baseline and robustness estimations**

| Estimator | FE | RE | PA | FE | RE | PA |
|---|---|---|---|---|---|---|
| GROWTH | -0.074*** | -0.081*** | -0.077*** | -0.073*** | -0.081*** | -0.077*** |
|  | (0.025) | (0.023) | (0.020) | (0.025) | (0.023) | (0.020) |
| GDPPC | -0.388 | -0.122 | -0.066 | -0.346 | -0.118 | -0.065 |
|  | (0.702) | (0.313) | (0.256) | (0.705) | (0.317) | (0.254) |
| CLAIM-PRIV | -0.002 | -0.001 | -0.001 | -0.003 | -0.001 | -0.001 |
|  | (0.002) | (0.001) | (0.001) | (0.002) | (0.002) | (0.001) |
| M2toRES | -0.209 | -0.213 | -0.212 | -0.221 | -0.221 | -0.216 |
|  | (0.225) | (0.225) | (0.183) | (0.226) | (0.226) | (0.182) |
| CPRIVET | 0.616** | 0.379 | 0.288 | 0.621** | 0.388 | 0.293 |
|  | (0.302) | (0.242) | (0.276) | (0.303) | (0.244) | (0.278) |
| GOV | 0.132 | -0.120 | -0.074 | 0.157 | -0.107 | -0.069 |
|  | (0.518) | (0.409) | (0.507) | (0.521) | (0.413) | (0.509) |
| INF | 2.810*** | 2.365*** | 1.767*** | 2.983*** | 2.451*** | 1.795*** |
|  | (0.850) | (0.697) | (0.494) | (0.860) | (0.724) | (0.499) |
| POLI | 0.127 | 0.167* | 0.174** | 0.129 | 0.167* | 0.174** |
|  | (0.101) | (0.086) | (0.087) | (0.101) | (0.086) | (0.086) |
| **FDIL** | **-1.743**** | **-1.515**** | **-1.187**** |  |  |  |
|  | (0.709) | (0.684) | (0.492) |  |  |  |
| **FDIT** |  |  |  | **-2.329**** | **-1.957**** | **-1.564**** |
|  |  |  |  | (0.797) | (0.758) | (0.563) |
| Constant |  | -15.115*** | -11.914*** |  | -15.579*** | -12.046*** |
|  |  | (4.219) | (2.964) |  | (4.365) | (2.965) |
| *Observations* | 783 | 1,499 | 1,499 | 783 | 1,499 | 1,499 |
| *Wald Test Statistic* | 44.69 | 38.67 | 51.75 | 47.40 | 39.59 | 52.52 |
| *Log Likelihood* | -198.3 | -291.3 |  | -197 | -290.3 |  |
| *Likelihood Ratio Test* | 44.69 | 31.89 |  | 47.40 | 32.75 |  |

Yearly data. Standard errors are presented below the corresponding coefficient. Marginal effects and the coefficients of the constant are reported. Symbols *, ** and *** indicate statistical significance at the 10%, 5%, and 1% levels, respectively.

## IV. Conclusions

This study shows that FDI not only have a positive direct effect on long-term growth in developing countries, but also increase the latter by reducing the recessionary effect resulting from a banking crisis. Even more, they reduce its occurrence. Consequently, it is recommendable for policymakers in developing countries to dispel the "false evidence" that has emerged since the 2008 international financial turmoil, according to which financial integration is an absolute synonym of crises.

Word count: 1993 words